# Sucralose Interaction with Protein Structures

*Nimesh Shukla[1], Enrico Pomarico[3], Cody J. S. Hecht[2], Erika A. Taylor[2], Majed Chergui[3] and Christina M. Othon[1,4]*

[1]Department of Physics, Wesleyan University, Middletown CT 06457 USA

[2]Department of Chemistry, Wesleyan University, Middletown CT 06457 USA

[3] Laboratoire de Spectroscopie Ultrarapide (LSU) and Lausanne Centre for Ultrafast Science (LACUS), École Polytechnique Fédérale de Lausanne, ISIC, FSB, CH-1015 Lausanne, Switzerland

[4] Molecular Biophysics Program, Wesleyan University, Middletown CT 06457 USA

## ABSTRACT

Sucralose is a commonly employed artificial sweetener that appears to destabilize protein native structures. This is in direct contrast to the bio-preservative nature of its natural counterpart, sucrose, which enhances the stability of biomolecules against environmental stress. We have further explored the molecular interactions of sucralose as compared to sucrose to illuminate the origin of the differences in their bio-preservative efficacy. We show that the mode of interactions of sucralose and sucrose in bulk solution differ subtly using hydration dynamics measurement and computational simulation. Sucralose does not appear to disturb the native state of proteins for moderate concentrations (<0.2 M) at room temperature. However, as the concentration increases, or in the thermally stressed state, sucralose appears to differ in its interactions with protein leading to the reduction of native state stability. This difference in interaction appears weak. We explored the difference in the preferential exclusion model using time-resolved spectroscopic techniques and observed that both molecules appear to be effective reducers of bulk hydration dynamics. However, the chlorination of sucralose appears to slightly enhance the hydrophobicity of the molecule, which reduces the preferential exclusion of sucralose from the protein-water interface. The weak interaction of sucralose with hydrophobic pockets on the protein surface differs from the behavior of sucrose. We experimentally followed up upon the extent of this weak interaction using isothermal titration calorimetry (ITC) measurements. We propose this as a possible origin for the difference in their bio-preservative properties.

## PICTORIAL ABSTRACT

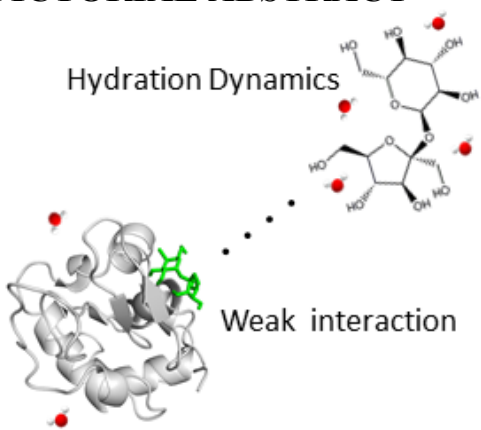

# INTRODUCTION

Sucrose is a naturally occurring small molecular osmolyte used to regulate the stability of solvated protein structures against environmental stress. Disaccharides such as sucrose and trehalose are capable of protecting biological organisms from diverse physical stresses including cryogenic storage [1, 2], elevated temperature [3, 4] dehydration [5], and excess salinity [6]; which has resulted in the widespread use of disaccharides in the cosmetic, food, and pharmaceutical industries. The biopreservation properties of disaccharide osmolytes can be attributed to their water structuring capabilities through preferential exclusion from the protein-water interface [7, 8]. We previously demonstrated [9] that the halogenation of sucrose, for the production of the artificial sweetener sucralose, resulted in the reduction of the biopreservation efficacy of the co-solute. The structure of each molecule is presented in schematic 1. In fact, while the native structure of various proteins was not impacted by the presence of sucralose for moderate concentrations at room temperatures, the stability against thermal stress was dramatically reduced in the presence of sucralose.

Sucrose protects the native state of protein structures against thermal denaturation in a concentration dependent manner. We found that the melting temperature of both Staph Nuclease and Bovine Serum Albumin increased linearly with sucrose concentration over the range of 0-0.5M [9]. Others have reported similar enhancement over a larger concentration range [10, 11]. Conversely, we found that sucralose strongly decreased the melting temperature of these model protein systems in a concentration dependent fashion. It appeared that the reduction of the biopreservation efficacy may have resulted from an alteration in the electrostatic properties of this molecule. In this work, we follow up on the impact halogenation has on the biopreservation efficacy of these molecules by modeling changes in the preferential exclusion of these molecules from the protein-water interface and investigated the water structuring capabilities of these two co-solutes.

The nature of the biomolecular interaction of sucralose has implications for understanding its bioavailability and its overall accumulation in living systems [12-20]. By measuring the water structuring capabilities of these two co-solutes we learn about the subtle attributes which can influence the efficacy of a solute as a bioprotective kosmotrope, and the degree to which two structurally similar molecules can modify the dynamics of bulk hydration dynamics. We have previously observed the change in hydration around halogenated biomolecules [21]. In that work, we observed that the fluorination of the natural amino acid leucine resulted in the dramatic reduction of interfacial solvation dynamics. Therefore, chemical modification of organic soluble molecules can dramatically impact the behavior of water within the surrounding solvation layer.

We have previously demonstrated that sucrose is capable of altering the dynamics of water far from the solvation layer of the disaccharide. We demonstrated a reduction of hydration dynamics at concentrations well below the overlap of the solvation layers between neighboring molecules [22]. Therefore, we expected that the chemical modification of sucrose might impact its ability to alter bulk hydration dynamics. In this study, we explored the effect of sucralose on bulk hydration dynamics and its impact on stability of proteins. We also explored the interaction of sucralose with four model protein systems and discussed its implication upon preferential exclusion mechanism of biopreservation. We accomplished this by using a combination of ultrafast optical spectroscopy and computational analysis on a variety of model systems.

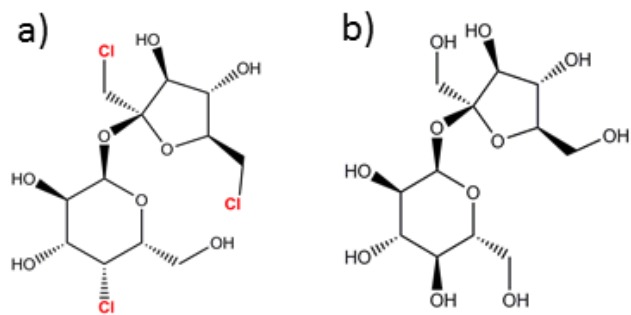

Schematic 1. a) Sucralose b) Sucrose.

**EXPERIMENTAL/SIMULATION METHODS**

*Sample Preparation.* Sucralose (98%) were purchased from Sigma-Aldrich, sucrose (99%) from Alfa Aesar and tryptophan (99%) from Acros Organics and were used without further purification. All the solutions were prepared from ultrapure 18 MΩ water. The concentration of tryptophan used in time resolved experiments was 3 mM. Stock sugar solutions were sonicated for 45 before preparation of target concentration. All the measurements were taken at room temperature (~21 °C).

*Steady State Spectroscopic Measurement.* Steady state fluorescence and UV-visible absorption spectroscopy were measured for all sugar concentrations. UV-visible absorption spectroscopy was measured on a Perkin Elmer spectrophotometer. Steady state fluorescence was measure on a Spex Fluoromax fluorometer using excitation at 295 nm with a 5 nm bandwidth. No change in the spectra and the total Stokes shift up to 0.1M of co-solute concentration, or lifetime of tryptophan up to 0.5M of co-solute concentration was observable. An example spectrum for 0.1M cosolute concentration and lifetime curve up to 0.5M co-solute concentration is given in the supplementary information (See Figure S1).

***Time-resolved Fluorescence Spectroscopy.*** The details of broadband fluorescence frequency up-conversion experimental setup are explained elsewhere [23]. In brief UV excitation pulses of wavelength 266 nm and pulse width of 70fs were used with repetition rate of 150 kHz. The emission fluorescence was filtered using a 300nm long pass filter to remove the residual excitation light and then an upconverted signal was obtained by mixing it with 800 nm gate beam. The frequency resolved upconverted signal was detected using a CCD camera equipped with a diffraction grating for dispersing the spectrum onto the detector. The broad-band measurements were done using a 250 μm BBO crystal, providing a time resolution of 250 fs as determined by fitting a measurement of the Raman response from the solvent. The sample solution was flown through a 0.2 mm thick quartz flow cell to avoid photodegradation. The sample flow and repetition rate were adjusted such as to achieve one shot per sample volume. The signal intensity and emission spectrum were monitored throughout the measurement to make sure no photodegradation or change in intensity occurs during the measurements. Before analyzing the time resolved fluorescence spectrum, Raman line of water was carefully removed, using a Gaussian fit of the up-converted Raman band in pure water. Significant caramelization and local heating can occur due to intense focusing of laser on small volume of a dense sample fluid; hence we restricted our study to low concentration of sugars at 0.1 M.

We fit the time resolved emission spectrums at each delay time using a modified log-normal function as described by Maroncelli and Fleming [24]. To calculate the solvation relaxation time constant, we extract the first moment from this lognormal fit. The first moment of the spectra were plotted as a function of delay time. This spectral relaxation was then fit using a three-exponential function convoluted with a Gaussian IRF. The value of Gaussian IRF and third time constant, representing lifetime of the probe, were kept fixed to 250fs and 3ns respectively.

***Simulation of Co-solute Interactions with Model Protein System.*** To investigate interaction of sucralose with proteins, computational ligand docking simulations were performed on four model protein systems. We chose reference enzyme systems where a bound substrate was also reported. This was done to provide a reference binding energy for a known substrate to calibrate the strength of the interactions of the cosolute and to verify proper simulation and docking of the substrate at the active site. The proteins used in this study along with their PDB code are hen egg-white lysozyme (1HEW) [25], Subtilisin DY (1BH6, a random mutant of subtilisin Carlsberg) [26], wild type Staphylococcal nuclease (4WOR) [27] and Thrombin Activatable Fibrinolysis Inhibitor (5LYD) [28]. For convenience, from now onwards we will refer to these proteins as Lysozyme, Subtilisin, SNase and Thrombin respectively. Docking simulations were performed using a publicly available ligand docking software Autodock Vina [29] along with AutoDockTools [30, 31]. Computation time on a high-performance computing cluster was provided by Wesleyan University. The structure of substrates of all proteins were downloaded from RCSB Protein Data Bank and structure of sucrose and sucralose were prepared using ChemDraw (PerkinElmer Informatics). Ligands conformations were energy minimized using MOPAC [32] before seeding the conformation in Autodock Vina. All the rotatable bonds were kept mobile. The metal ions reported in crystal structure of proteins were kept present for all docking simulations. Proteins structures were kept rigid and the value of exhaustiveness parameter was fixed to 100. A total of 10000 poses were generated for each ligand by running Autodock Vina 500 times, each time using a random seed conformation and a random iteration parameter. To ensure the validity of docking algorithm used by Autodock Vina, substrates were removed from crystal structure of proteins, seeded, and then docked to their respective proteins. All simulations of sucrose and sucralose were performed after making sure that substrate of each protein was docking correctly at its active site as reported in the crystal structure. For highlighting local hydrophobicity on protein surface a YRB color code scheme proposed by Hagemans *et. al* [33]. was produced using the python script provided in their work. Visualization, editing and printing of docking results were done using PyMol [34].

*Calorimetric Analysis.* Isothermal titration calorimetry was performed on a VP-ITC MicroCalorimeter (MicroCal, LLC). All measurements were done using an Origin based controlling software provided by MicroCal. A long equilibration time was provided to achieve baseline deviation within 0.02μCal/Sec. All samples were prepared in 0.05 M $NaH_2PO_4$ buffer solution at pH 7.5 with 10μM of $CaCl_2$. In titration cell ~1.5ml of protein sample containing ~84 μM of wild type Staphylococcal Nuclease in buffer was maintained at 25 °C and stirred at 307 rpm. A total of 55 injections of 5μL each of 0.5 M sucrose and sucralose solution in buffer were delivered over 10 seconds with 3.5 min of equilibration time in between the successive injections. Change in enthalpy (ΔH) curves as a function of molar ratio were obtained by integrating area under the raw ITC data. Two reference curves were measured, first by injecting 0.5M sugar solution in buffer and second by injecting buffer into the protein solution giving enthalpy of dilution of sugars and enthalpy of dilution of protein sample respectively.

## RESULTS AND DISCUSSION

*Slowdown of Hydration Dynamics by Sucralose.* To investigate the effect chlorination may have on the biomolecular interactions of sucrose we measured interaction of both co-solutes with the intrinsically fluorescent biomolecule, tryptophan. We measured steady state fluorescence and emission spectra for sucrose and sucralose solutions containing 3 mM of tryptophan. No change in the spectra, the total Stokes shift, or the lifetime of tryptophan was observable for co-solute concentration up to 0.5M. The lack of any spectral shift in the presence of either sucrose or sucralose indicates that there is no direct interaction between the sweeteners and the probe molecule. This is consistent with sucrose's role as a non-specific osmolyte that only stabilizes proteins through a solvent mediated, indirect interaction [8]. For sucralose, this implies that any interaction between the probe and the co-solute must occur during the non-equilibrium interactions introduced by the excited state of the probe. The lack of apparent changes in either the absorption or emission spectra suggest that with the interactions are long range and transient, and we infer from this that sucralose is not directly bound to tryptophan. Tryptophan was used as the fluorescent probe in our broadband fluorescence frequency up-conversion experiment to measure changes in solvent dynamics through the solvatochromic Stokes shift [35, 36].

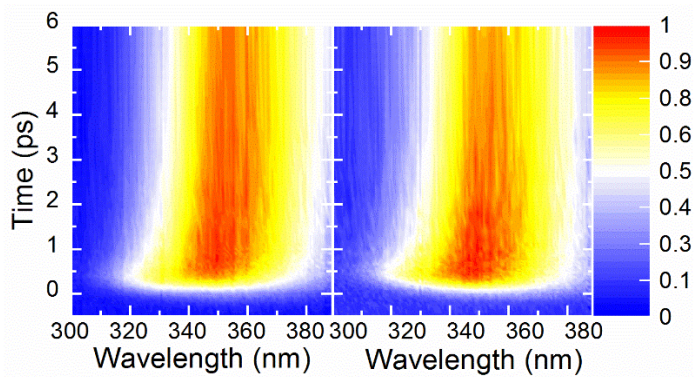

Figure 1: Normalized spectral intensity relaxation of tryptophan at 0.1 M concentration of co-solutes. A Sucralose spectrum is presented in the left panel and Sucrose in the right panel.

We measured time-resolved changes in hydration dynamics around the solvated tryptophan due to presence of sucrose and sucralose in the bulk. The full spectral response for both sucrose and sucralose at 0.1 M is shown in Figure 1. The dynamic Stokes shift in time resolved emission spectrums is directly correlated with solvation dynamics of the fluorophore molecule tryptophan.

We observed a solvation time constant of 2.8 ± 0.6 ps for sucralose and 1.5 ± 0.4 ps for sucrose, see Figure 2. Previously Bräm *et.al.*[36] have reported the solvation time constant of 1.02 ± 0.12 ps for tryptophan in pure water. Solvation time constant in presence of sucralose is ~86% larger compared to sucrose, this suggest that sucralose is more effective in reduction of bulk hydration dynamics than sucrose.

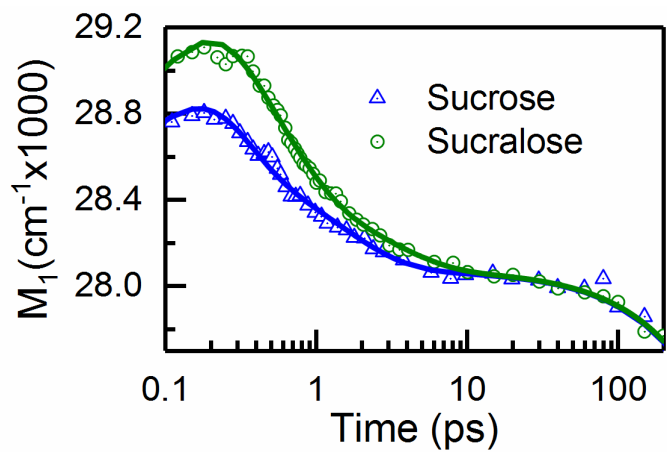

Figure 2: Spectral relaxation of tryptophan plotted as a function of time for 0.1 M sucralose and 0.1 M sucrose solutions; the error bars are comparable to the size of the data points. A three-exponential fit convoluted with Gaussian IRF accurately represents the relaxation as shown by solid lines.

The chlorination of sucrose to derive sucralose appears to have reduced its solubility. We found that the solubility limit of sucralose is approximately 0.6 M, however, sucrose is soluble in water up to ~6 M. Increase in hydrophobicity of proteins (BSA, gelatin and ovalbumin) upon chlorination have previously been reported by Seguchi in 1985 [37]. He found that three residues namely tyrosine, lysine and cystine became more hydrophobic upon chlorination and no change was observed for all other residues. Increase in hydrophobicity upon chlorination of promazine, perazine and perphenazine analogues has also been reported by Gerebtzoff *et. al*. in 2004 [38]. It is possible that the chlorination of sucrose has altered both its water structuring capabilities as well as its interactions with the protein interface.

The increase in melting point as a function of disaccharide concentration for multiple protein systems has been presented previously [10, 11]. We also previously reported this behavior for BSA and staphylococcal nuclease in presence of sucrose, sucralose and trehalose [9, 22]. We observed that for sucrose and trehalose the increase in melting temperature could be directly correlated with slowdown in bulk water hydration dynamics [22], which supports a preferential exclusion mechanism of biopreservation by disaccharides. Preferential exclusion implies that there is no direct interaction between disaccharides and proteins (or biomolecule). The addition of disaccharides to bulk water sequesters water molecules away from the protein, decreasing its hydrated radius and increasing its compactness and consequently stability [8]. Although sucralose modifies the bulk water dynamics much more effectively than sucrose, counter intuitive to preferential exclusion model for biopreservation we found it to be destabilizing in nature for proteins [9]. It is possible that this difference arises from the degree to which these two co-solutes are preferentially excluded from protein surface. Sucralose may interact weakly with proteins at high concentrations or within a thermally stressed environment. This would render the larger water structuring capability of sucralose ineffective in promoting the stability of biomolecules. As it seems the solubility of sucralose has been reduced as compared to sucrose, we hypothesized that sucralose could have slightly higher affinity for binding near hydrophobic interfaces on protein surfaces.

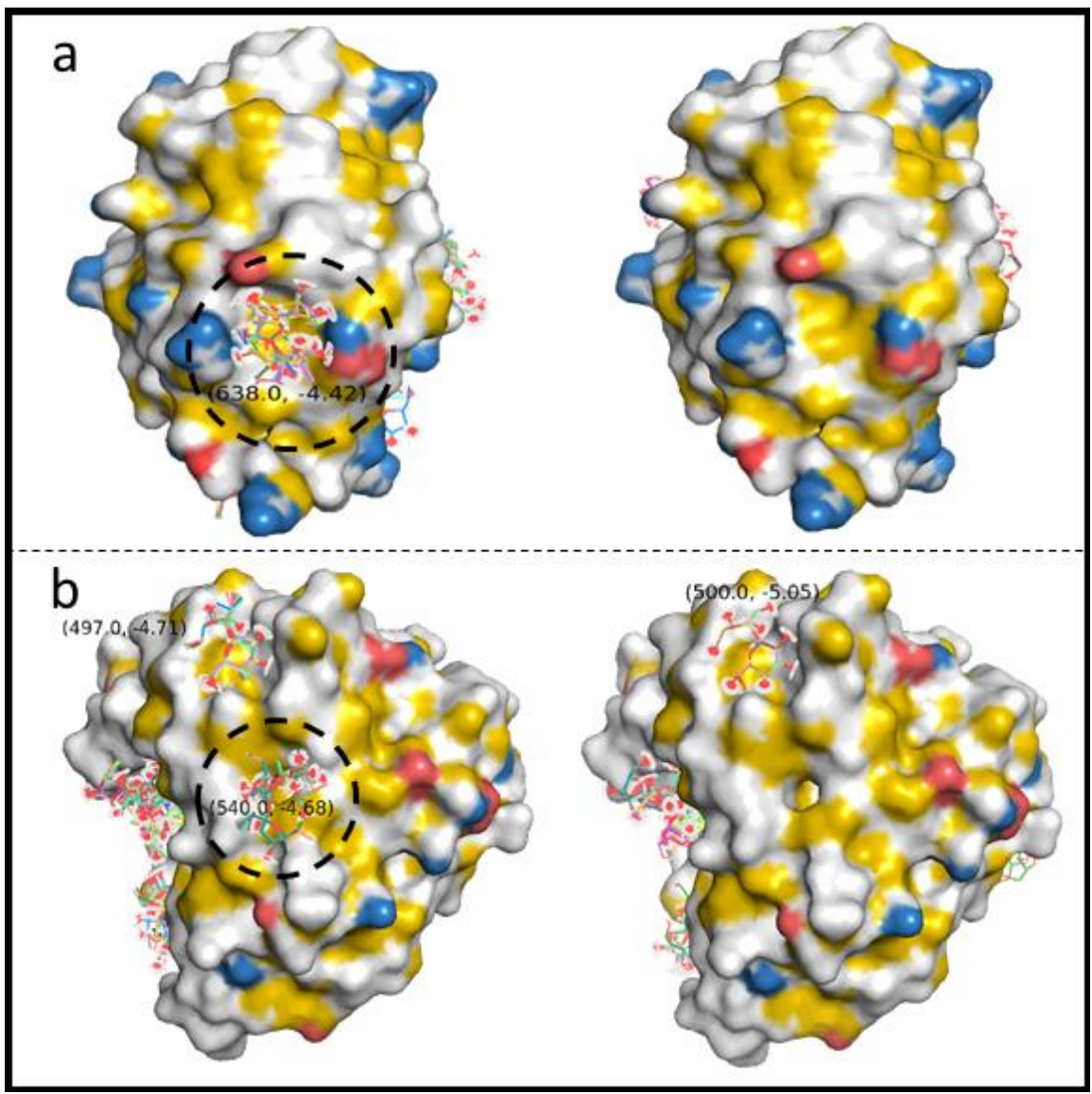

Figure 3: Binding conformations of sucralose (left) and sucrose (right) when substrate was left bound at the active site for two model enzymes a) Lysozyme b) Subtilisin. The unique interaction of sucralose with hydrophobic pocket on protein surfaces is highlighted using dotted black circle.

***Interaction of sucralose with hydrophobic surfaces of proteins.*** To investigate this hypothesis, ligand docking simulations were performed on four model enzyme systems. At first all proteins were docked using their substrate as ligand. This provided a reference energetic binding level to compare to the co-solute interactions. The proper binding of the ligand to the binding site reported in the crystal structure also provides a good test to ensure the simulation is operating as expected. In order to observe any low frequency events, we generated 10,000 docking poses for each substrate, the mean and least binding energies are reported in table-S1. We found that the orientation of least energy conformation of each substrate matches very well with the orientation of substrate reported in the crystal structure, see Figure S2. We also observed that majority of substrate conformations actually ended up binding with the active site (100% for Lysozyme and SNase, 98.35% for Subtilisin, 73.1% for thrombin), see Figure S2. This observation built our confidence in the docking algorithm used by Autodock Vina.

Since it is possible that docking of sucrose and sucralose to proteins may be dominated by their affinity to bind at the active site, and in order to explore entire surface of the protein, simulations were performed in two ways: a) when the bound substrate at the active site is left in a bound state observed in the crystal structure and; b) when the substrate is deleted from the active site. We docked sucralose and sucrose to all model proteins systems 10,000 times, the mean and least binding energies for all conformations are reported in table S1. We observed that, generally, sucrose and sucralose bound to the proteins with a lower calculated affinity than did their substrates. Additionally, we observed no significant difference in either the least or mean binding energy of sucralose compared to sucrose for all proteins. This observation suggests that the difference between sucralose and sucrose is very subtle, probably due to some specific property of the binding sites. Upon further analysis, we found that there are few binding sites unique for sucralose where sucrose does not bind or binds very weakly, see Figure 3, Figure S3, S4, S5, and S6. The hydrophobic pockets on surface of proteins are highlighted in yellow using the YRB color code scheme proposed by Hagemans et.al. As shown in Figure 3.a, 638 (~6.4%) conformations of sucralose bind at the hydrophobic pocket on the surface of lysozyme while no conformation of sucrose binds at this site. Similarly, for subtilisin (Figure 3.b), 540 (~5.4%) conformations of sucralose binding at the hydrophobic pocket while no conformation of sucrose binds at this site. This effect appears weak but is reproducible in successive docking simulations performed using a random seed conformation and a random iteration parameter. The consistency of the hydrophobic nature of the unique sucralose binding sites across different model systems, and the fact that these conformations represent a sizable fraction of the non-active site conformations, make it statistically unlikely that these are random binding events. This affinity could change the preferential exclusion of sucralose and could result in transient, low-energy weak interaction with proteins at moderate to high co-solute concentration. To further explore this hypothesis, we performed Isothermal Titration Calorimetry on one protein staphylococcal nuclease.

*Weak interaction of sucralose with proteins.* Isothermal titration calorimetry is a well-established technique for studying protein ligand interaction. It provides change in enthalpy of reaction upon binding of ligand as a function of molar ratio which could further be used to obtain various thermodynamic quantities such as entropy, Gibbs free energy and binding constant. We measured the change in enthalpy of system as function of molar ratio of staphylococcal nuclease and sugars. The raw data and enthalpy curves are shown in Figure 3.

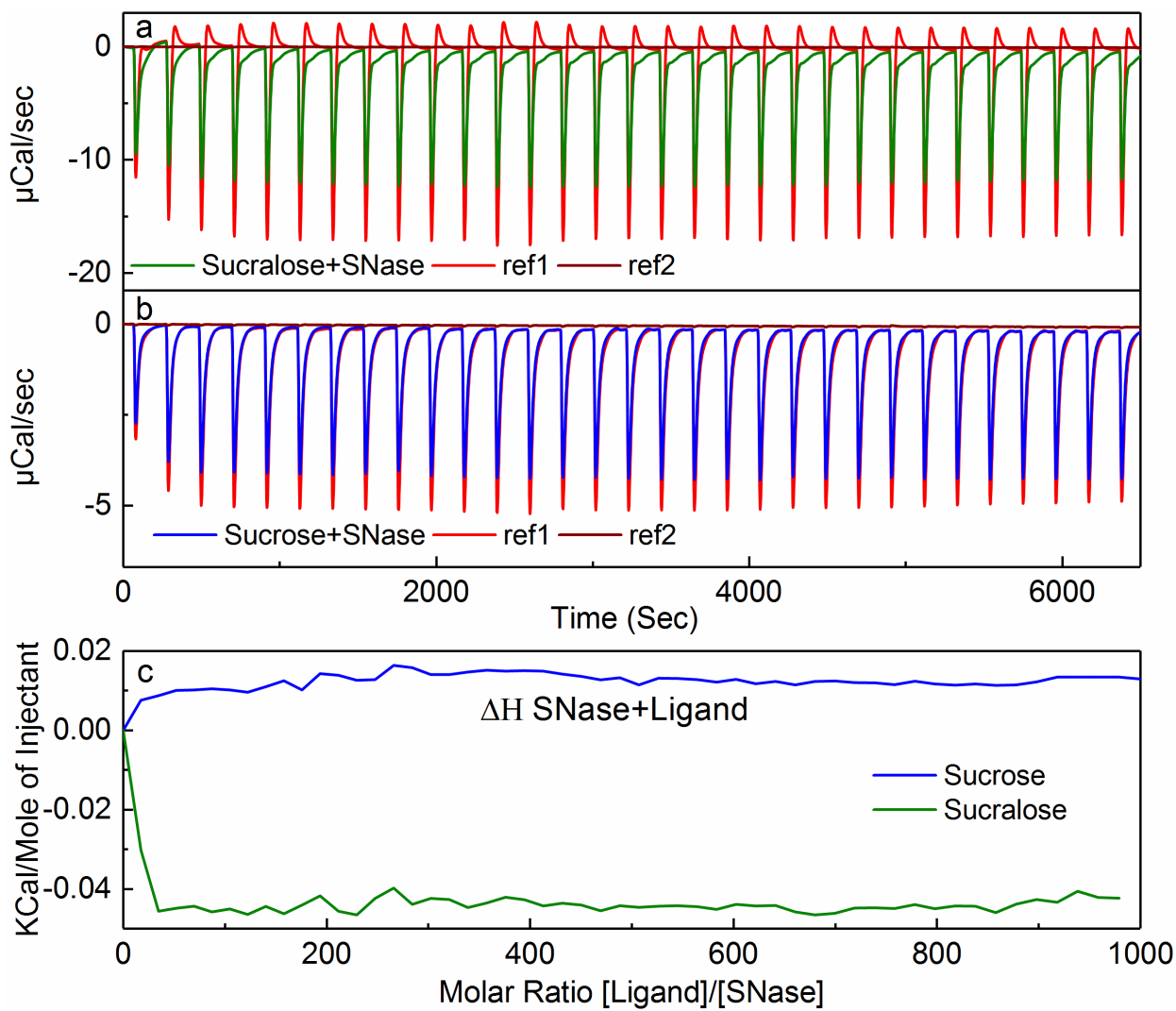

Figure 4: a) Heat evolved upon titration of SNase with injection of sucralose (Sucralose+SNase), with injection of buffer (ref2) and titration of Buffer with injection of sucralose (ref1). b) Heat evolved upon titration of SNase with injection of sucrose (Sucrose+SNase), with injection of buffer (ref2) and titration of Buffer with injection of sucrose (ref1). c) Change in enthalpy as a function of molar ratio obtained by subtracting the integration of ref1 and ref2 from Sugar+Snase for sucrose and sucralose.

The raw data obtained upon titration of SNase by injecting sugars is shown in Figure 4.a and 4.b. Upon calculating changes in enthalpy of interaction of sugars and SNase, we found it to be very small, of the order of few µCal/Mol hence in order to isolate the interaction it becomes important to subtract the enthalpy of dilution of sugars and protein as they are also similar order of magnitude. The ref1 curve measures the enthalpy of dilution of the sugars obtained by titrating buffer solution with injection of sugars, upon which the concentration changes from ~0.5M to ~0.084M. This substantial change in concentration of sugars results into comparable change in enthalpy for ref 1 (see Figure 4.a and 4.b) and Sugars+SNase curve. As can be seen from the ref 1 curve for sucrose and sucralose both shows a negative exothermic peak indicative of release in heat energy upon dilution. Sucralose also shows a small positive endothermic peak indicating absorption of heat energy upon dilution of the sucralose solution. We have observed that sucralose differs from sucrose in both its bulk water structuring capabilities and its electrostatic dipole moment [9] therefore, this small difference in the enthalpy of dilution is not surprising. The ref2 curve measures enthalpy of dilution of protein obtained by titrating SNase solution with injection of buffer, upon which concentration of SNase get diluted from ~84µM to ~70µM. As this is a very small change in concentration of SNase, the ref 2 (see Figure 4.a and 4.b) curve is very similar to the baseline and indicates no noticeable change in enthalpy. The true interaction of SNase with sugars is obtained by subtracting the integration of ref1 and ref2 from Sugar+SNase curve and plotted in Figure 4.c as a function of molar ratio. In Figure 4.c, the positive but negligible value ~0.01Kcal/Mol of change in enthalpy of interaction of sucrose with SNase suggests that either sucrose has no affinity to interact with SNase or has very small tendency to repel SNase, which is consistent with preferential exclusion of sucrose from protein surface. Sucralose on the other hand has a very small negative change in enthalpy indicative of either no interaction or perhaps a weak interaction with proteins. The very small amplitude of change in enthalpy for sucralose ~0.045KCal/Mol rules out any hydrogen bond formation among sucralose and protein as it requires three orders of magnitude more binding energy ~10Kcal/Mol. This suggests that the interaction between sucralose and the protein must be weak dipole-dipole type of interaction. In highly stressed environments and at high concentrations of sucralose this weak interaction could cause sucralose to bind with proteins and alter its folding dynamics which in turn destabilizes protein structures.

This weak interaction could alter the structural stability of proteins as it needs to explore a new energy conformation phase space, which may or may not be stable at elevated temperatures. Our previous work [9] indicated that sucralose did not significantly alter the native state conformation of model proteins at room temperature or for concentrations below 0.2 M. Therefore, we anticipated that the reduction in melting point was due to a subtle shift in the interactions of sucralose with the protein. A small shift in the hydrophobic effect or weak dipolar interactions would be consistent with the available data. This subtle shift is also consistent with the apparent contradiction in the water structuring capabilities of this molecule. While sucralose is more effective than sucrose at reducing hydration dynamics in the bulk, it is not an effective bio-preservative. While solvation of many molecules might result in modified hydration dynamics, it is only those that are strongly excluded from the protein interface that would enhance the protein stability through the preferential exclusion model.

Our data provides insight into the origin of the bio-preservative properties of compatible osmolytes. It is important to note that the quality of the simulation results is only as good as the ability of such simulations to model hydrophobic interactions. The reliable prediction of the role of hydrophobicity in protein-ligand binding is not without controversy [39]. However, we feel that the consistency of the nature of the unique binding sites of sucralose provide a strong argument for the origin of this very subtle change in protein interaction. The result is consistent with its change in water solubility as well. We feel that the sucralose-sucrose comparison could be informative for the further development of hydrophobic binding of ligands. Furthermore, the change in protein interaction is valuable for food scientists investigating the bioavailability, stability, and processing of sucralose food items.

## CONCLUSION

We explored the effect of sucralose on bulk hydration dynamics using ultrafast up-conversion spectroscopy and demonstrated 180% slowdown in presence of sucralose, which is ~3.5 times more than sucrose at the same concentration. We found that although sucralose is a superior dynamic reducer of bulk hydration dynamics that it is destabilizing in nature for proteins. Our docking simulation suggest that sucralose has a slightly larger tendency to bind with hydrophobic pockets on protein surfaces. This could result in a weak interaction of sucralose with proteins which in turn destabilizes protein structures.

## ACKNOWLEDGEMENTS

We thank Dr. Adam Willard and Sucheol Shin for their useful discussions. This work has been partially supported by the Wesleyan Grants in Support of Scholarship and by the NCCR:MUST of the Swiss National Science Foundation.

## REFERENCES

1. Somme L. and Meier T., *Cold tolerance in tardigrada from dronning-maud-land, antarctica.* Polar Biology, 1995. **15**(3): p. 221-224.

2. Storey K.B. and Storey J.M., *Frozen and alive.* Scientific American, 1990. **263**(6): p. 92-97.

3. Benaroudj N., Lee D.H., and Goldberg A.L., *Trehalose accumulation during cellular stress protects cells and cellular proteins from damage by oxygen radicals.* Journal of Biological Chemistry, 2001. **276**(26): p. 24261-24267. **DOI:** 10.1074/jbc.M101487200

4. Sola-Penna M. and Meyer-Fernandes J.R., *Stabilization against thermal inactivation promoted by sugars on enzyme structure and function: Why is trehalose more effective than other sugars?* Archives of Biochemistry and Biophysics, 1998. **360**(1): p. 10-14. **DOI:** 10.1006/abbi.1998.0906

5. Crowe J.H., *Anhydrobiosis: An unsolved problem.* American Naturalist, 1971. **105**(946): p. 563-573. **DOI:** 10.1086/282745

6. Wright J.C., Westh P., and Ramlov H., *Cryptobiosis in tardigrada.* Biological Reviews of the Cambridge Philosophical Society, 1992. **67**(1): p. 1-29.

7. Choi Y., et al., *Molecular dynamics simulations of trehalose as a 'dynamic reducer' for solvent water molecules in the hydration shell.* Carbohydrate Research, 2006. **341**(8): p. 1020-1028. **DOI:** 10.1016/j.carres.2006.02.032

8. Jain N.K. and Roy I., *Effect of trehalose on protein structure.* Protein Sci., 2009. **18**(1): p. 24-36. **DOI:** 10.1002/pro.3

9. Chen L., et al., *Sucralose destabilization of protein structure.* The Journal of Physical Chemistry Letters, 2015. **6**(8): p. 1441-1446. **DOI:** 10.1021/acs.jpclett.5b00442

10. Kumar A., Attri P., and Venkatesu P., *Trehalose protects urea-induced unfolding of alpha-chymotrypsin.* Int. J. Biol. Macromol., 2010. **47**(4): p. 540-5. **DOI:** 10.1016/j.ijbiomac.2010.07.013

11. Kaushik J.K. and Bhat R., *Why is trehalose an exceptional protein stabilizer? An analysis of the thermal stability of proteins in the presence of the compatible osmolyte trehalose.* J Biol Chem, 2003. **278**(29): p. 26458-65. **DOI:** 10.1074/jbc.M300815200

12. Lange F.T., Scheurer M., and Brauch H.J., *Artificial sweeteners-a recently recognized class of emerging environmental contaminants: A review.* Analytical and Bioanalytical Chemistry, 2012. **403**(9): p. 2503-2518. **DOI:** 10.1007/s00216-012-5892-z

13. Mawhinney D.B., et al., *Artificial sweetener sucralose in u.S. Drinking water systems.* Environmental Science & Technology, 2011. **45**(20): p. 8716-8722. **DOI:** 10.1021/es202404c

14. Mead R.N., et al., *Occurrence of the artificial sweetener sucralose in coastal and marine waters of the united states.* Marine Chemistry, 2009. **116**(1-4): p. 13-17. **DOI:** 10.1016/j.marchem.2009.09.005

15. Tollefsen K.E., Nizzetto L., and Huggett D.B., *Presence, fate and effects of the intense sweetener sucralose in the aquatic environment.* Science of the Total Environment, 2012. **438**: p. 510-516. **DOI:** 10.1016/j.scitotenv.2012.08.060

16. Wiklund A.K.E., et al., *Sucralose induces biochemical responses in daphnia magna.* Plos One, 2014. **9**(4). **DOI:** 10.1371/journal.pone.0092771

17. Wiklund A.K.E., et al., *Sucralose - an ecotoxicological challenger?* Chemosphere, 2012. **86**(1): p. 50-55. **DOI:** 10.1016/j.chemosphere.2011.08.049

18. Labare M.P. and Alexander M., *Microbial cometabolism of sucralose, a chlorinated disaccharide, in environmental-samples.* Applied Microbiology and Biotechnology, 1994. **42**(1): p. 173-178.

19. Schiffman S.S. and Abou-Donia M.B., *Sucralose revisited: Rebuttal of two papers about splenda safety.* Regulatory Toxicology and Pharmacology, 2012. **63**(3): p. 505-508. **DOI:** 10.1016/j.yrtph.2012.05.002

20. Schiffman S.S. and Rother K.I., *Sucralose, a synthetic organochlorine sweetener: Overview of biological issues.* Journal of Toxicology and Environmental Health-Part B-Critical Reviews, 2013. **16**(7): p. 399-451. **DOI:** 10.1080/10937404.2013.842523

21. Kwon O.H., et al., *Hydration dynamics at fluorinated protein surfaces.* Proc Natl Acad Sci U S A, 2010. **107**(40): p. 17101-6. **DOI:** 10.1073/pnas.1011569107

22. Shukla N., et al., *Retardation of bulk water dynamics by disaccharide osmolytes.* The Journal of Physical Chemistry B, 2016. **120**(35): p. 9477-9483. **DOI:** 10.1021/acs.jpcb.6b07751

23. Cannizzo A., et al., *Femtosecond fluorescence upconversion setup with broadband detection in the ultraviolet.* Optics Letters, 2007. **32**(24): p. 3555. **DOI:** 10.1364/ol.32.003555

24. Maroncelli M. and Fleming G.R., *Picosecond solvation dynamics of coumarin 153: The importance of molecular aspects of solvation.* The Journal of Chemical Physics, 1987. **86**(11): p. 6221. **DOI:** 10.1063/1.452460

25. Cheetham J.C., Artymiuk P.J., and Phillips D.C., *Refinement of an enzyme complex with inhibitor bound at partial occupancy.* Journal of Molecular Biology, 1992. **224**(3): p. 613-628. **DOI:** 10.1016/0022-2836(92)90548-X

26. Eschenburg S., et al., *Crystal structure of subtilisin dy, a random mutant of subtilisin carlsberg.* European Journal of Biochemistry, 1998. **257**(2): p. 309-318. **DOI:** 10.1046/j.1432-1327.1998.2570309.x

27. Wall M.E., Ealick S.E., and Gruner S.M., *Three-dimensional diffuse x-ray scattering from crystals of staphylococcal nuclease.* Proceedings of the National Academy of Sciences of the United States of America, 1997. **94**(12): p. 6180-6184.

28. Halland N., et al., *Sulfamide as zinc binding motif in small molecule inhibitors of activated thrombin activatable fibrinolysis inhibitor (tafia).* Journal of Medicinal Chemistry, 2016. **59**(20): p. 9567-9573. **DOI:** 10.1021/acs.jmedchem.6b01276

29. Trott O. and Olson A.J., *Autodock vina: Improving the speed and accuracy of docking with a new scoring function, efficient optimization, and multithreading.* Journal of Computational Chemistry, 2010. **31**(2): p. 455-461. **DOI:** 10.1002/jcc.21334

30. Morris G.M., et al., *Autodock4 and autodocktools4: Automated docking with selective receptor flexibility.* Journal of computational chemistry, 2009. **30**(16): p. 2785-2791. **DOI:** 10.1002/jcc.21256

31. Sanner M.F., *Python: A programming language for software integration and development.* Journal of Molecular Graphics & Modelling, 1999. **17**(1): p. 57-61.

32. Stewart J.J.P., *Mopac.* 2016, Stewart Computational Chemistry.

33. Hagemans D., et al., *A script to highlight hydrophobicity and charge on protein surfaces.* Frontiers in Molecular Biosciences, 2015. **2**: p. 56. **DOI:** 10.3389/fmolb.2015.00056

34. Schrodinger, LLC, *The pymol molecular graphics system, version 1.8.* 2015.

35. Qin Y., et al., *Validation of response function construction and probing heterogeneous protein hydration by intrinsic tryptophan.* The Journal of Physical Chemistry B, 2012. **116**(45): p. 13320-13330. **DOI:** 10.1021/jp305118n

36. Bram O., et al., *Relaxation dynamics of tryptophan in water: A uv fluorescence up-conversion and molecular dynamics study.* J. Phys. Chem. A, 2010. **114**(34): p. 9034-42. **DOI:** 10.1021/jp101778u

37. Seguchi M., *Model experiments on hydrophobicity of chlorinated starch and hydrophobicity of chlorinated surface protein.* Cereal Chemistry, 1985. **62**(3): p. 166-169.

38. Gerebtzoff G., et al., *Halogenation of drugs enhances membrane binding and permeation.* Chembiochem, 2004. **5**(5): p. 676-84. **DOI:** 10.1002/cbic.200400017

39. Snyder P.W., et al., *Mechanism of the hydrophobic effect in the biomolecular recognition of arylsulfonamides by carbonic anhydrase.* Proceedings of the National Academy of Sciences of the United States of America, 2011. **108**(44): p. 17889-17894. **DOI:** 10.1073/pnas.1114107108

# Data in Brief Article

**Title**: Data for sucralose interaction with protein structure.

*Nimesh Shukla[1], Enrico Pomarico[3], Cody J. S. Hecht[2], Erika A. Taylor[2], Majed Chergui[3] and Christina M. Othon[1,4]*

[1]Department of Physics, Wesleyan University, Middletown CT 06457 USA

[2]Department of Chemistry, Wesleyan University, Middletown CT 06457 USA

[3] Laboratoire de Spectroscopie Ultrarapide (LSU) and Lausanne Centre for Ultrafast Science (LACUS), École Polytechnique Fédérale de Lausanne, ISIC, FSB, CH-1015 Lausanne, Switzerland

[4] Molecular Biophysics Program, Wesleyan University, Middletown CT 06457 USA

**Contact email**: cothon@wesleyan.edu

**Abstract**

Destabilizing nature of sucralose for proteins in contrast with its counterpart sucrose has been demonstrated earlier[1]. To study the effect of sucralose on bulk hydration dynamics a probe fluorophore tryptophan was used. In this article, we present results of steady state measurements of absorption/emission spectrums and life time of tryptophan at various concentration of sucralose and sucrose. In addition, we also present results from ligand docking simulations performed on four model protein systems (Lysozyme, Subtilisin, SNase and Thrombin). Least and mean binding energies of substrate and sugars with for proteins are presented in the Table S1. For verification of ligand docking algorithm used by Autodock VINA the comparisons between docked conformation of substrate and experimentally obtained conformation from crystal structure is presented in Figure

S2. Ligand docking simulations were then performed to revel molecular level difference in interaction of sucralose and sucrose with proteins. Figure S3-S6 presents comparison between docked conformations of sucralose and sucrose with four model protein systems.

**Specifications Table** [*Please fill in right-hand column of the table below.*]

| Subject area | *Physics, biophysics, biochemistry, physical chemistry* |
|---|---|
| More specific subject area | *Hydration dynamics, Biopreservation* |
| Type of data | *Table, text file, graph, figure* |
| How data was acquired | *Spex Fluoromax fluorometer, Autodock VINA, Pymol* |
| Data format | *Analyzed.* |
| Experimental factors | *Not applicable.* |
| Experimental features | *Steady state fluorescence and UV-visible absorption spectroscopy were measured for all sugar concentrations. UV-visible absorption spectroscopy was measured on a Perkin Elmer spectrophotometer. Steady state fluorescence was measure on a Spex Fluoromax fluorometer using excitation at 295 nm with a 5 nm bandwidth.* |
| Data source location | *Not applicable.* |
| Data accessibility | *Data attached with this article.* |
| Related research article | *Not applicable.* |

**Value of the Data**

- The data represents steady state absorption/emission and lifetime of tryptophan in presence of sucralose and sucrose.
- The data also presents verification of ligand docking algorithm used by Autodock VINA for four model protein systems.
- The data visualize unique or weak interaction of sucralose with four model protein systems simulated using Autodock VINA.
- This data will provide complimentary support to arguments presented in the main article.

**Data**

The data presented in this article are experimental and computational results indicating no direct interaction of sugars with probe fluorophore tryptophan and a unique but weak interaction of sucralose with four model protein systems.

***Steady State Absorption and Emission Spectrum.*** The steady state absorption and emission spectrum for tryptophan in sucrose and sucralose solutions are indistinguishable, see figure S1. The total stokes shift and fluorescence lifetime were identical for all concentrations investigated in this study indicating that neither of the co-solutes is directly interacting with the probe molecule on the timescale of the lifetime of the molecule.

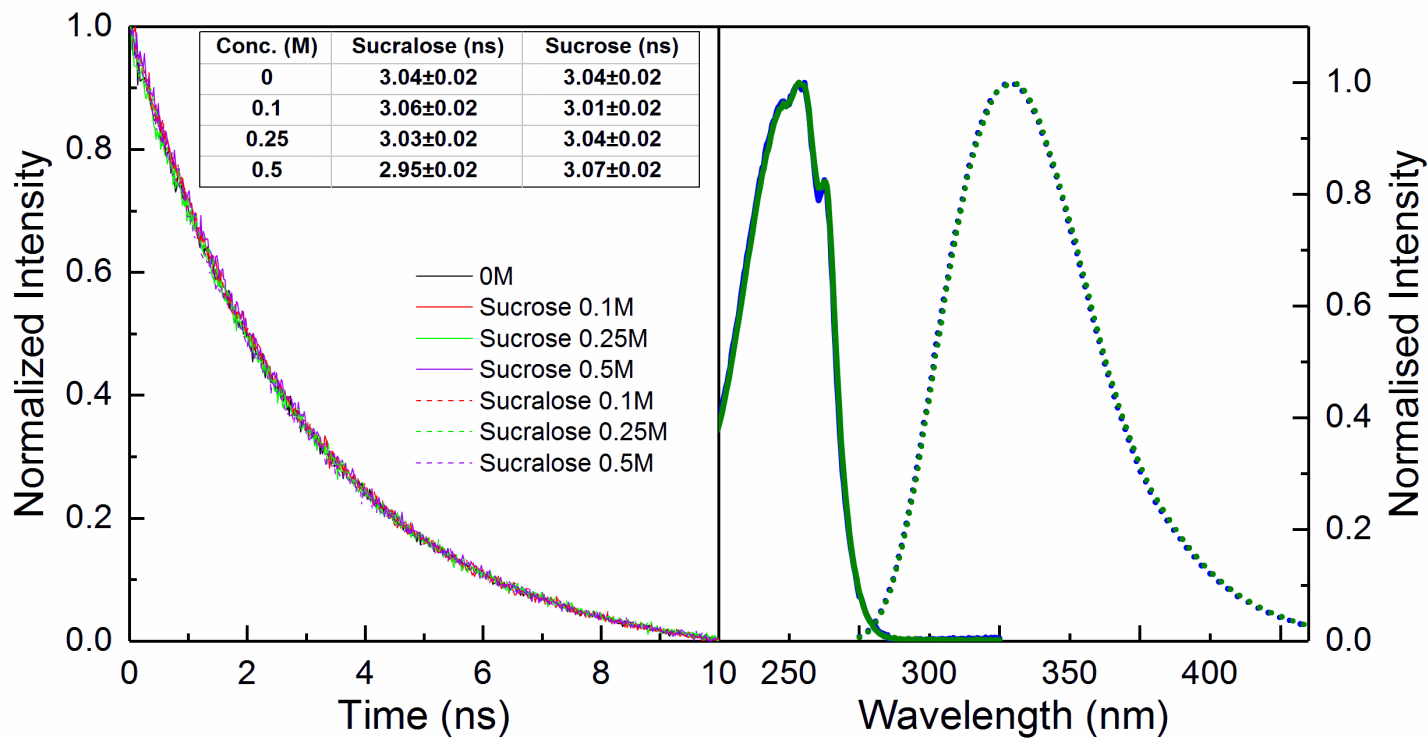

**Figure S1**. **Left,** Lifetime curve up to 0.5M concentrations of co-solutes. **Right,** Steady-state absorption (solid line) and emission spectra (dotted line) for 0.1 M sucrose (blue) and 0.1 M sucralose (green) solution, for 3 mM tryptophan sample.

| Protein → | Lysozyme(1HEW) | | | | Subtilisin(1BH6) | | | | SNase (4WOR) | | | | Thrombin(5LYD) | | | |
|---|---|---|---|---|---|---|---|---|---|---|---|---|---|---|---|---|
| Substrate → | Absent | | Present | | Absent | | Present | | Absent | | Present | | Absent | | Present | |
| Ligand↓ | L | M | L | M | L | M | L | M | L | M | L | M | L | M | L | M |
| Substrate | 6.2 | 5.5 | 5.4 | 4.8 | 8.4 | 7.2 | 7.5 | 6.1 | 8.0 | 7.0 | 6.3 | 5.7 | 7.0 | 5.3 | 5.5 | 4.9 |
| Sucralose | 5.9 | 5.2 | 5.2 | 4.6 | 5.7 | 5.0 | 5.3 | 4.7 | 6.0 | 4.9 | 5.7 | 4.9 | 5.6 | 5.0 | 6.1 | 5.2 |
| Sucrose | 6.3 | 5.6 | 5.6 | 5.1 | 5.8 | 5.3 | 5.5 | 4.9 | 6.1 | 5.3 | 5.2 | 4.7 | 5.8 | 5.1 | 5.5 | 5.2 |

**Table S1:** Least (L) and Mean (M) binding energies of ligands with their respective proteins. The RCSB protein database ID number for the crystal structure is shown in parenthesis next to the enzyme name. Binding energies are reported in the unit of -Kcal/Mol.

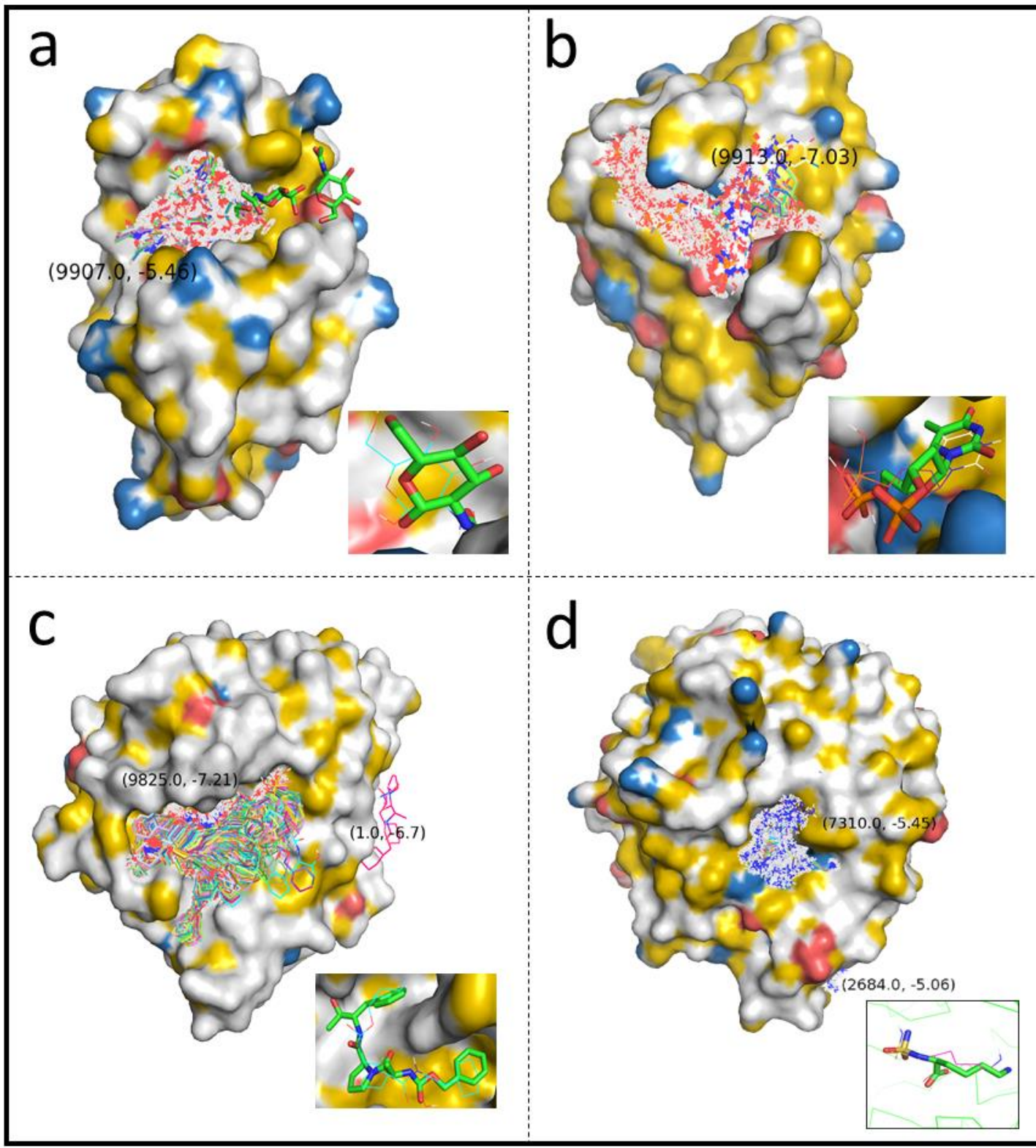

**Figure S2:** Image of all 10000 conformations for each substrate docked to its respective proteins a) Lysozyme(1HEW) b) SNase(4WOR) c) Subtilisin(1BH6) d) Thrombin(5LYD). Inset showing

zoom out view of substrate at the active site with stick model as the reported crystal structure and wireframe as docked conformation of least binding energy. Total number of conformations in each cluster along with its average binding energy is also shown in brackets.

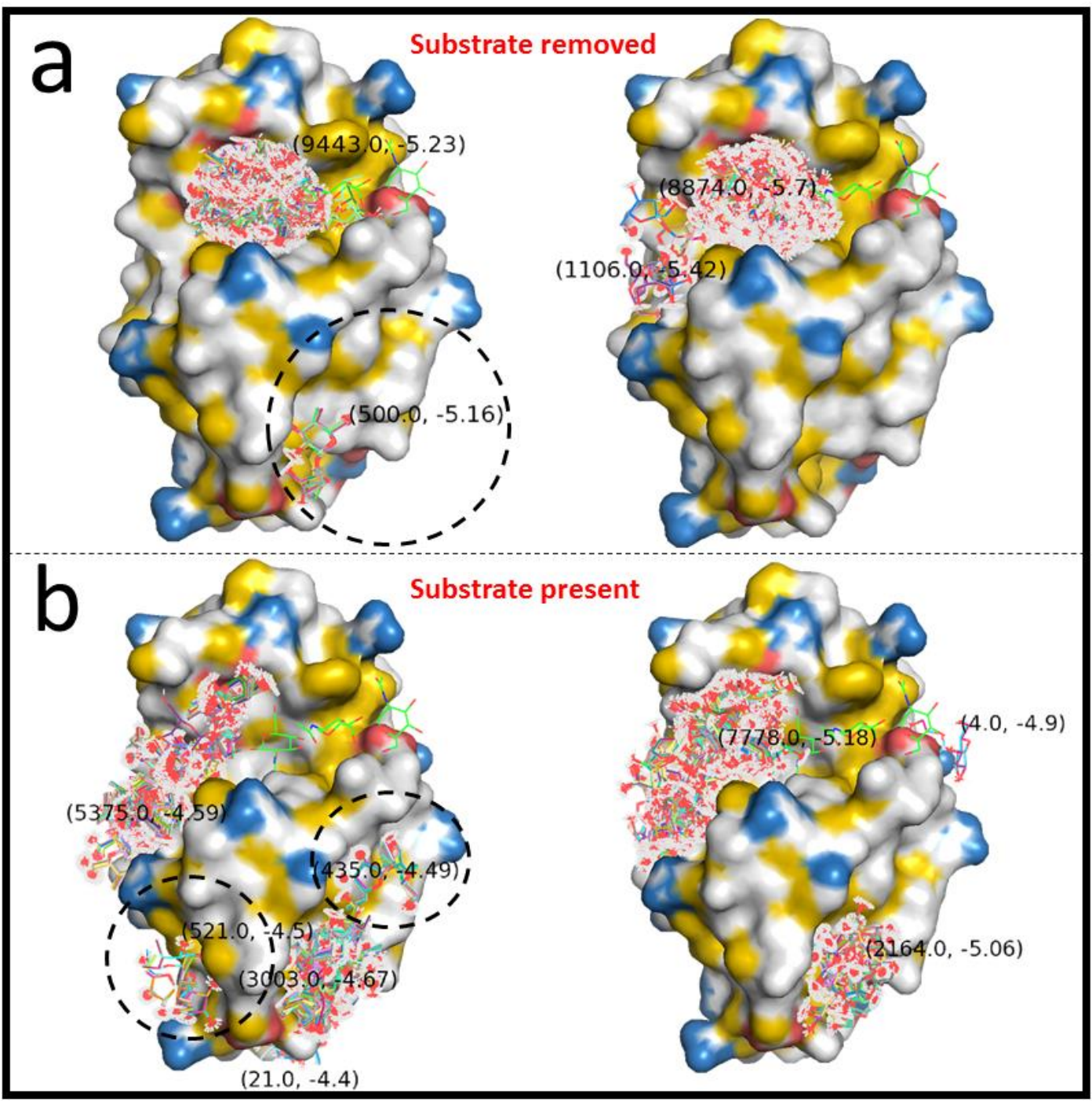

**Figure S3:** Docked conformations of sucralose (left) and sucrose binding (right) for Lysozyme (1HEW) a) when substrate is removed from the active site b) when substrate is left bound at the active site. The unique/stronger interaction of sucralose compared to sucrose with hydrophobic

parts on proteins is highlighted using dotted black circle. Total number of conformations in each cluster along with its average binding energy is also shown in brackets.

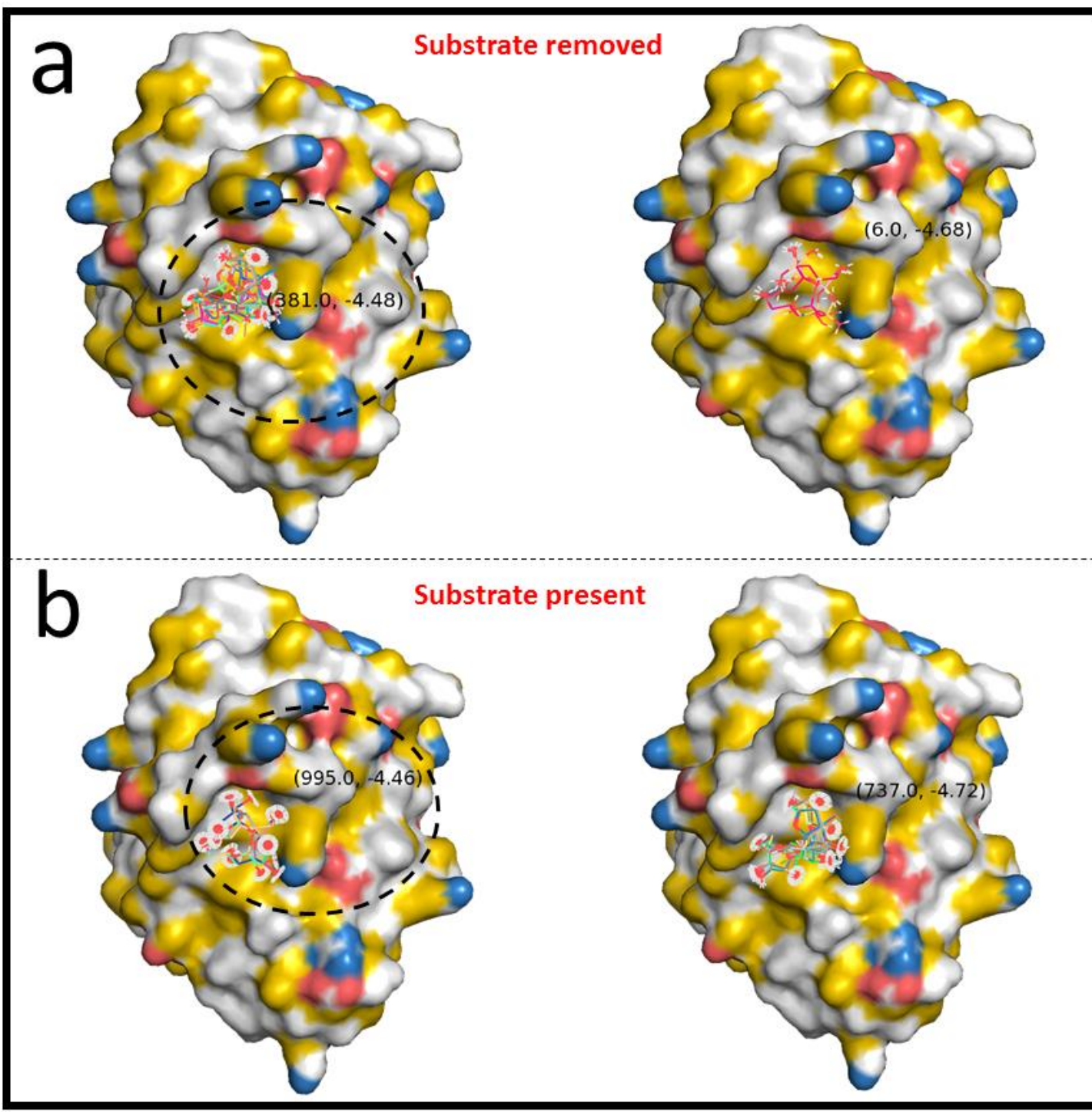

**Figure S4:** Docked conformations of sucralose (left) and sucrose (right) for SNase (4WOR) when a) Substrate is removed from the active site b) Substrate is left bound at the active site. The unique/stronger interaction of sucralose compared to sucrose with hydrophobic parts on proteins is highlighted using dotted black. Total number of conformations in each cluster along with its average binding energy is also shown in brackets.

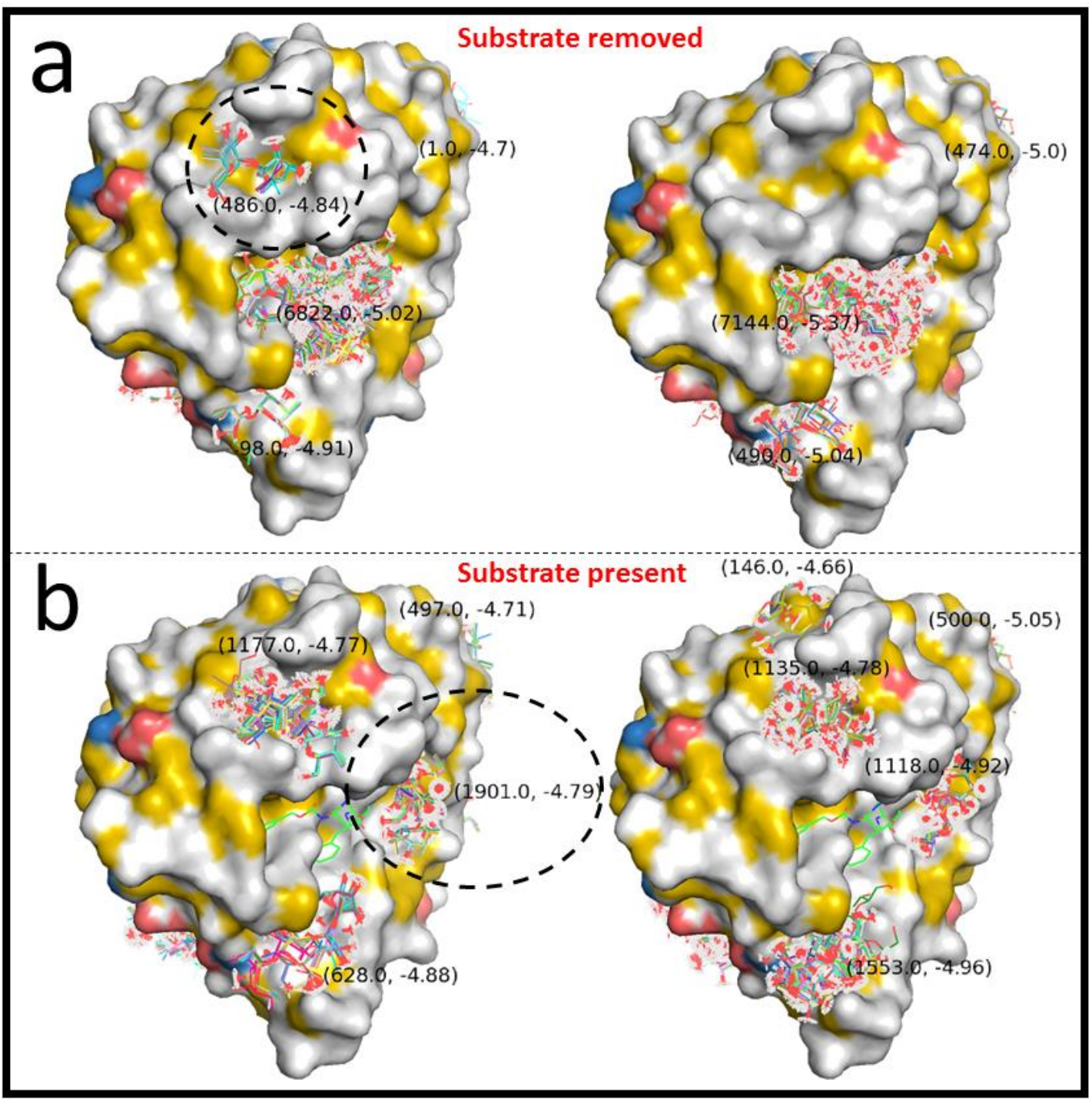

**Figure S5:** Docked conformations of sucralose (left) and sucrose (right) for Subtilisin (1BH6) when a) Substrate is removed from the active site b) Substrate is left bound at the active site. The unique/stronger interaction of sucralose compared to sucrose with hydrophobic parts on proteins

is highlighted using dotted black. Total number of conformations in each cluster along with its average binding energy is also shown in brackets.

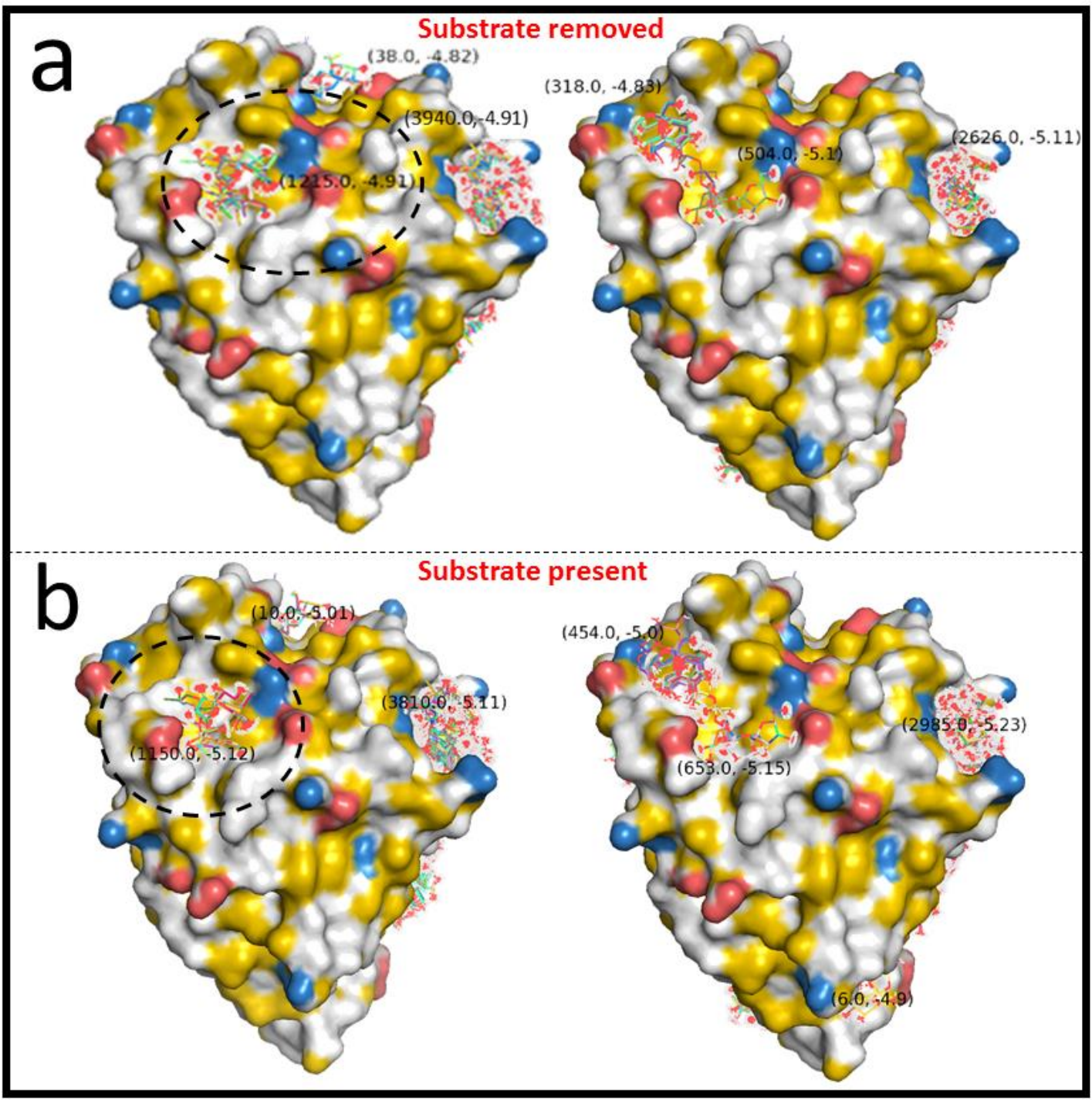

**Figure S6:** Docked conformations of sucralose (left) and sucrose (right) for Thrombin (5LYD) when a) Substrate is removed from the active site b) Substrate is left bound at the active site. The unique/stronger interaction of sucralose compared to sucrose with hydrophobic parts on proteins is highlighted using dotted black. Total number of conformations in each cluster along with its average binding energy is also shown in brackets.

**References:**

1. Chen L., et al., *Sucralose destabilization of protein structure.* The Journal of Physical Chemistry Letters, 2015. **6**(8): p. 1441-1446. **DOI:** 10.1021/acs.jpclett.5b00442